\documentclass[pdflatex,sn-mathphys-num]{sn-jnl}
\usepackage[utf8]{inputenc}
\usepackage{graphicx}%
\usepackage{multirow}%
\usepackage{amsmath,amssymb,amsfonts}%
\usepackage{amsthm}%
\usepackage{mathrsfs}%
\usepackage[title]{appendix}%
\usepackage{xcolor}%
\usepackage{textcomp}%
\usepackage{manyfoot}%
\usepackage{booktabs}%
\usepackage{algorithm}%
\usepackage{algorithmicx}%
\usepackage{algpseudocode}%
\usepackage{listings}%
\usepackage{graphicx}
\usepackage{float} 
\usepackage[version=4]{mhchem}
\begin{document}
\title[UHDR, the FLASH effect, and \ce{H2O2} yields: Do Experiments and Simulations really disagree?]{Ultra-High Dose-Rates, the FLASH Effect, and Hydrogen Peroxide Yields:\\ Do Experiments and Simulations Really Disagree?}
\author*[1]{\fnm{Marc Benjamin} \sur{Hahn}*}\email{marc-benjamin.hahn@fu-berlin.de}
\affil*[1]{\orgdiv{Institut f\"ur Chemie}, \orgname{Universit\"at Potsdam}, \orgaddress{\street{Karl-Liebknecht-Str. 24-25}, \postcode{14476}, \city{Potsdam},  \country{Germany}}}
\abstract{
Radiation chemistry of model systems irradiated with ultra-high dose-rates (UHDR) is the key to
obtain a mechanistic understanding of the observed sparing of healthy tissue. This sparing is called the FLASH effect. It is
envisioned to be used for more efficient treatment of cancer by \emph{FLASH radiotherapy}. However, it seems that even the most simple model systems, namely water irradiated with varying dose-rates, pose a challenge.
This became evident recently, as differences within measured and predicted
hydrogen peroxide (H$_{2}$O$_{2}$) yields (g-values) for exposure
of liquid samples to conventional dose-rates and UHDR were reported. Many of
the recently reported values contradict older experimental studies
and current Monte-Carlo simulations (MCS).\\ 
In the present work, we aim to identify possible underlying reasons
of these discrepancies and propose ways to overcome this issue. Hereby
a short review of recent and classical literature concerning experimental
and simulational studies is performed. The studies cover different
radiation sources, from gamma rays, high-energy electrons, heavy particles,
such as protons and ions, with low and high linear energy transfer
(LET), and samples of hypoxic and oxygenated water, with and without
added cosolutes such as bovine-serum albumine (BSA). Results are compared
in terms of additional experimental parameters, such as solvent, sample
container and analysis methods used to determine the respective g-values
of H$_{2}$O$_{2}$. Similarly the parameter governing the outcome
of the MCS by the step-by-step (SBS) approach, or the independent-reaction
time (IRT) method are discussed. Here, UHDR induced modification
of the radical-radical interaction and dynamics, not governed by diffusion
processes, may cause problems. Approaches to test these different models are highlighted
to allow progress:  by making the step from a purely descriptive discourse
of the effects observed, towards testable models, which should clarify
the reasons of how and why such a disagreement came to light in the
first place.
}
\keywords{Radiation damage, Radiotherapy, FLASH, UHDR, LDR, Oxygen, FLASH radiotherapy, ROS, Geant4-DNA, Topas-nBio, Hydrogen peroxide, PCR, \ce{H2O2}}
\maketitle
\section{Introduction}
The sparing of normal tissue by ultra-high dose-rates (UHDR, above 40Gy/s) compared to low, conventional dose rates (LDR, below 2Gy/s) 
is known as the FLASH effect. This effect is envisioned to make the treatment of radio resistant tumors with radiotherapy more efficient
by enhancing the therapeutic window.\cite{vozeninmechanisms2025}
Despite that the effects of UHDR are already being studied in living organisms, and first preclinical studies and treatments are performed,\cite{bourhistreatment2019}
the underlying radiobiological mechanisms are far from being understood.
Here, one of the main question is: How does a physical effect, acting on a nanosecond to picosecond timescale, lead to a differential radiation response in healthy and cancerous tissue? \\
It has been established that the inelastic scattering cross sections for various tissues remain constant with respect to the dose rate.
Consequently, the observed normal tissue sparing effects must be triggered by subsequent radiation chemical mechanisms. These mechanisms influence
the cellular radiation response and potentially affect the organism on organ level, or the immune response as a whole.\\
However, in order to comprehend the intricate intra- and intercellular mechanisms that dictate the fate of irradiated tissue, it is imperative to attain a comprehensive understanding of the initial step, that exhibits a dependence on the dose rate (DR): radiation chemistry.
Or as pointed out previously in other words: \emph{``Radiation chemistry comes before radiation biology''} \cite{oneillradiation2009} - and
if one get's the first wrong, the second will be wrong alike. \cite{wardmanradiotherapy2020}
Therefore, we will focus in the following mainly on the chemical processes triggered by water radiolysis - and meanwhile identify some problems, which should motivate us to spend a little more time thinking about the physico-chemical stage during radiation exposure in model systems - before diving into the full complexity of biology.\\
\subsection*{The problem}
Since the discovery of the FLASH effect two main 'chemical' hypotheses were put forward to explain the differential radiation response in cells by UHDR. 
One assumes fast radical-radical recombination between different radiation tracks during short timescales, while the second proposed variation in the oxygen yield. 
Especially the latter process seems to be a natural candidate to lead to a difference in cell survival, due to the varying oxygen levels present in healthy and tumor cells.\cite{mckeowndefining2014}
The often found hypoxic (low) oxygen levels in many tumors make them more radiation resistant than healthy tissue with normoxic conditions.\cite{vonsonntagfreeradicalinduced2006}
However, recent experiments showed, that the DR dependent oxygen depletion seemed to be an unlikely cause of the sparing effects due to insufficient oxygen consumption to make a biological difference.\cite{jansendoes2021,thomasproton2024}
In contrast, the DR dependence of radical recombinations is a well known process studied for decades by pulse radiolysis techniques.
\cite{andersonradiation1962,schwarzdetermination1962,adamsreactions1965}\\
However, recent studies performed with proton or electron beams, which delivered clinically relevant dose and DRs, lead to a contradiction between experimentally measured radical yields and the computationally predicted g-values when studying UHDR. Most of these studies focused on measuring radical yields by chemical assays in water or buffered solutions. 
Hereby hydrogen peroxide was the most common endpoint. 
\ce{H2O2} was chosen due to it's long lifetimes in these \emph{in-vitro} systems, and the easy possibility to quantify it by fluorescence based assays without involving more complex, time resolved spectroscopy.
The g-value of \ce{H2O2} depends on the yields and reactions of the primary radicals produced by water radiolysis. These radicals react with each other, and compete for reactions with dissolved oxygen and other cosolutes. Surprisingly, the relative \ce{H2O2} yields were found to be lower, when transitioning from LDR to UHDR exposure, which stands in stark contrast to results from older radiolysis studies, the yields predicted by solving the related reaction kinetics, or performing Monte-Carlo simulations (MCS).\cite{zhanganalysis2024,d-kondointegrated2023}\\
Multi-scale MCS are envisioned to play a vital role in accessing properties involved in the radiation response that are difficult to access by experimental methods alone. 
Thereby they have to bridge the gap between physico-chemical processes and the biological response, which stretch over multiple orders of time and length scales.\cite{hahnaccessing2023}
To be able to do so, it is imperative to get the radiation chemistry right, and resolve the different outcomes between experiment and simulation, and to clarify which minimal requisites have to be met to extrapolate results from model system with reduced complexity, to a more realistic ones. 
Therefore the experimental conditions will be examined with respect to their differences in terms of settings, assays, bunch and pulse parameters and their outcome.
\section{The state of the art}
To resolve the presented inconsistencies between reported data of dose-rate dependent hydrogen peroxide yiels, we will briefly summarize the most recent work and the classical studies of the field: Recently, Montay-Gruel \emph{et al.}\cite{montay-gruellongterm2019} observed higher hydrogen peroxide yields in water with 4\% O$_{2}$, pH7 under conventional DR than under UHDR. 
Irradiations were performed with 6MeV electrons in 1-2$\,$\textmu s bunches, varying DRs, in Eppendorf PCR tubes made from polyproprylen and analyzed with the \emph{Amplex Red} assay. Kacem \emph{et al.}\cite{kacemcomparing2022} observed similar behaviour when comparing 5.5$\,$MeV electron and 235$\,$MeV protons in water with 4\% O$_{2}$ in water irradiated in PCR tubes and analyzed with \emph{Amplex Red}. The reduction of the g-value of H$_{2}$O$_{2}$ for electrons was about 34\,\% lower when increasing the DR from 0.1\,Gy/s to above
1400\,Gy/s. For proton irradiations the g-value decreased by 18\% when increasing the DR from 0.9\,Gy/s to 1260\,Gy/s. Exposure to 235\,keV LDR xrays with 0.035\,Gy/s showed higher g-values than all corresponding electron and proton exposures. Sunnerberg \emph{et al.}\cite{sunnerbergmean2023}
applied electron beams with 9-10$\,$MeV energy, 3.5$\,$\textmu s bunch width, and with varying instantaneous and mean DR in the range of (10$^{2}$-10$^{6}$)$\,$Gy/s and (0.14-1500)$\,$Gy/s, respectively. 
Experiments were done in aqueous solutions with added bovine serum albumin (BSA) to provide a protein microenvironment within polystyrene cuvettes. Under these conditions, the highest dose-rate of 1500$\,$Gy/s showed an about 3.3 times lower yield of H$_{2}$O$_{2}$ per Gy than the the lowest DR of 0.14$\,$Gy/s. After further analysis, they concluded that the ``Mean dose rate in ultra-high dose rate electron irradiation is a significant predictor for O$_{2}$ consumption and H$_{2}$O$_{2}$ yield''.\cite{sunnerbergmean2023} Thomas \emph{et al.}\cite{thomasproton2024}
performed experiments with 250$\,$MeV protons a bunch width of 2$\,$ns, an average DR of 10$\,$Gy/s and 80$\,$Gy/s with corresponding instantaneous dose rates of 68$\,$Gy/s and 550$\,$Gy/s. Additionally irradiations with 10$\,$MeV electrons a bunch width of 4$\,$ms, an average DR of 0.1$\,$Gy/s and (115-660$)\,$$\,$Gy/s with corresponding instantaneous
DR of 108$\,$Gy/s and around 10$^{6}$$\,$Gy/s were performed. The experiments were performed in water with and without additional BSA. 
Samples were irradiated in PCR tubes and for the H$_{2}$O$_{2}$ assay \emph{Amplex Red} was applied. Under all these conditions a higher H$_{2}$O$_{2}$ yield was observed with conventional DRs than
for UHDR conditions. The relative difference was highest for the electron irradiated samples containing BSA. Zhang \emph{et al.}\cite{zhanganalysis2024}
performed their experiments with 430$\,$MeV/u carbon ions (50$\,$Gy/s, 0.1$\,$Gy/s), 9$\,$MeV electrons (0.6$\,$Gy/s and 600$\,$Gy/s) and 200$\,$kV xrays (DR 0.1$\,$Gy/s and 10$\,$Gy/s). 
They were carried out under ``real hypoxic'' (1$\,$\% O2, 0.1$\,$\% CO2), hypoxic (1$\,$\% O$_{2}$, 5$\,$\% CO$_{2}$) and normoxic (21$\,$\% O$_{2}$) conditions. Some of the measurements were conducted with varying bunch and bunch-train structure, with and without bubbling of N$_{2}$O to convert hydrated electrons into hydroxyl radicals, as well as additional electron scavengers. 
The experiments were performed in PCR tubes made from polypropylene with ultrapure water in the pH range between 6-7, and with the \emph{Amplex UltraRed} assay for measuring the hydrogen peroxide yields. For the majority of settings they found a higher H$_{2}$O$_{2}$ yield for conventional DRs compared to UHDR.
As soon as hydrated electron scavenger were present, this difference vanished, with the exception of 9$\,$MeV electrons under hypoxic conditions, where the scavenging was achieved by dissolved N$_{2}$O in the solution. 
Zhang \emph{et al.} proposed a model based on variations in DR dependent intertrack recombinations of the various radical species based on their different diffusion constants. They assumed that the
ratio of the reactions represented by the equations 
\begin{equation}
\ce{e_{aq}^{-} + ^{.}OH -> ^{-}OH\qquad}(k=3\cdot10^{10}\,{\textstyle\,M^{-1}s^{-1}}),\label{eq:e-OH}
\end{equation}
\begin{equation}
\ce{^{.}OH + ^{-}OH -> ^{-.}O + H2O}\qquad(k=1.2\cdot10^{10}\,{\textstyle\,M^{-1}s^{-1}}),
\end{equation}
and
\begin{equation}
\ce{^{.}OH+^{.}OH -> H_{2}O_{2}}\quad(k=0.55\cdot10^{10}\,{\textstyle\,M^{-1}s^{-1}}),\label{eq:Oh-Oh}
\end{equation}
changes in a DR dependent manner, since the intertrack yields during the non-homogeneous chemical phase changes for high DR. 
This was attributed to the lower diffusion constant of the hydroxyl radical (2.2$\cdot$10\textsuperscript{-9}m\textsuperscript{2}/s) compared to that of the hydrated electron (4.9\,$\cdot$10\textsuperscript{-9}m\textsuperscript{2}/s) and hydroxide (5.3\,$\cdot$10\textsuperscript{-9}m\textsuperscript{2}/s), which favor these species to react with other partners from different tracks already during the inhomogeneous chemical stage. 
This was supposed to become relevant for UHDR conditions where these tracks are on average closer in space and time. This modified competition between the species was proposed to lead to a depletion of \ce{^.OH-^.OH} recombinations and a lower \ce{H2O2} yield.\\
In contrast to these recent studies, previous work showed opposing results. 
The experiments by Anderson and Hart \cite{andersonradiation1962} were performed with with 15$\,$MeV electrons bunches of 1.4$\,$\textmu s at UHDR pulse length in water at different oxygen levels in syringes, and the \ce{H2O2} yields were determined by the triiodide method and titration. 
The comparision between a LDR Co$^{60}$and the UHDR electron source at 1$\,$mM O$_{2}$ revealed a 60$\,\%$ higher g-value for H$_{2}$O$_{2}$ production under UHDR than under LDR. They attributed
the DR dependent increase of the hydrogen peroxide yield to the dominance of \ce{^{.}OH + ^{.}OH -> \ce{H2O2}} over the various other reactions
considered (\emph{e.g.} of \ce{^{.}OH} with \ce{HO2^.} and \ce{e_{aq}^{-}})
Similarly, a later study by Sehstedt and Rasmussen \cite{sehestedrate1968} was performed with 10$\,$MeV electrons involving single and multiple
microsecond bunches, with bunch doses in the order of Gy and average DRs between 5-50$\,$Gy/s, in a pH range of 0.46-6.75 in oxygenated 
water (1.2$\,$mM O$_{2}$) within ``Pyrex reaction cells'' made from borosilicate glass and in the presence of different scavengers. 
They also found an increase of \ce{H2O2} yields with the increase of DR at pH 6.75. From these studies it was concluded, that the competition between the reaction described in Eq.\,\ref{eq:e-OH} and Eq.\,\ref{eq:Oh-Oh} together with the following reactions
\begin{equation}
\ce{H^{.} + O2 ->HO^{.}\quad}(k=2.1\cdot10^{10}\mathrm{\,M^{-1}s^{-1}}),
\end{equation}
\begin{equation}
\ce{HO2^{.} + ^{.}OH ->H2O + O2\quad}(k=1\cdot10^{10}\mathrm{\,M^{-1}s^{-1}}),
\end{equation}
 and 
\begin{equation}
\ce{2 HO2^{.}\longrightarrow H2O2 + O2\quad}(k=6\cdot10^{5}\mathrm{\,M^{-1}s^{-1}\;at\:pH7}),
\end{equation}
govern the hydrogen peroxide yield, which is about 0.07\,\textmu M/Gy in unbuffered oxygenated water at pH7.\cite{wardmanradiationchemical2023,wardmanradiotherapy2020,andersonradiation1962,buxtoncritical1988} 
Various calculations and simulational studies on dose-rate effects showed similar results as the older works mentioned here, namely an increase in hydrogen peroxide yields with dose-rate when simulations where performed in pure water systems with a total dose higher than about 1$\,$Gy.\cite{d-kondointegrated2023,derksenmethod2023,shininvestigation2024,hahnsparing2025,wardmanradiotherapy2020,alanazicomputer2021}
This is not too surprising, since the simulational frameworks are validated\cite{ramosmendeztopasnbio2021,wardmanapproaches2022}
against these g-values, and based on the rate constants compiled in the reference tables from the literature, and references therein. \cite{buxtoncritical1988,vonsonntagfreeradicalinduced2006,vonsonntagchemical1987} \\
So far, the simulation study by \emph{Shin et al.}\cite{shininvestigation2024} provided the most comprehensive comparison between g-values from experimental data and simulations over a wide range of beam types
and parameters. This included a simulation of the settings based on
the aforementioned studies by Sehestedt \emph{et al.} \cite{sehestedrate1968}
and Kacem \emph{et al.} \cite{kacemcomparing2022}. Overall they found
a good agreement between simulations and experimental results, with
the exception of the LDR dataset of Kacem \emph{et al.} \cite{kacemcomparing2022},
where 235$\,$MeV protons in 4$\,$\% O$_{2}$ and pH7 at DR of 0.9$\,$Gy/s
were concerned. A similar disagreement between experiment and simulation
in the LDR regime was observed in the same study,\cite{shininvestigation2024}
when they simulated the DR dependent oxygen consumption as measured
by Jansen \emph{et al.}\cite{jansendoes2021}, while HDR results agreed
with each other. Possible reasons for these discrepancies will be
examined in the following.
\section{Experimental Answers?}
Generally the studies can be classified as showing either a higher
relative hydrogen peroxide yield for HDR ($\frac{g_{LDR}(H_{2}O_{2})}{g_{HDR}(H_{2}O_{2})}<1$)
or for LDR ($\frac{g_{LDR}(H_{2}O_{2})}{g_{HDR}(H_{2}O_{2})}>1$ ).
The 'older studies' and simulations fall into the former category,
while the more 'recent' studies all belong to the latter. To gain
an initial insight, it makes sense to take a look at their similarities
and differences.\\
All 'recent' studies were performed with a broad range of particle
types, beam energies, bunch structures, instantaneous DRs and mean
DRs. The differences instantaneous DRs and mean DRs are sometimes
not clearly reported and accounted for which makes a direct interpretation
of such studies difficult. For example, two irradiations performed
with the same mean DR (averaged over a macroscopic 'second' timescale),
can have very different instantaneous DRs (DR within an isolated bunch/pulse
or respective bunch/pulsetrain), which lead obviously to very different
radical yields, since the related reactions happen on microsecond
and sub-microsecond timescales. \cite{schulerultrahigh2022} Nevertheless,
since the range of irradiation conditions used in the recent studies
was very broad, the overall results still showed similar tendencies for
the DR dependent g-values, which makes it unlikely that the reason
can be rooted in small variations within these parameters. In contrast,
common denominators of these studies are, that they all (1) used the
\emph{Amplex red} assay to determine the H$_{2}$O$_{2}$ yield, and
that these experiments were performed (2) in containers made out
of various types of plastic materials. These two reasons seem to be
the most obvious systematic differences to the 'older' studies and
simulations.
\subsection*{The simple solution: materials}
The latter may provide the most straight forward explanation: The
purity of solvents and container materials are of utmost importance
for reliable measurements in radiation chemistry, since even small
amounts of reactants or catalysts present, can change the reaction
dynamics and outcomes substantially. Therefore, the use of thoroughly
purified samples and solvents as well as extensively cleaned, inert
fused silica vessels are recommended for mechanistic studies, to avoid
contamination's.\cite{wardmanmechanisms2023} Thus, especially the
aforementioned plastic vessels may turn out to have unwanted side
effects, when studying the g-values of reactive oxygen species. For
example, it is well known that plastics like polyproprylene and polystyrene
can release oxygen into samples and were even found to alter the oxygen
enhancement ratio in cell cultures exposed to radiation.\cite{chapmanradiosensitization1971,chapmanfactors1970,kochrelease1972}
Similarly nanoplastic or microplastic \cite{braunsmart2021} could
be released from the container as well, and act as a scavenger in
the solution, resulting in the alteration of the hydroxyl radicals,
which would be able to completely modifying and influence the g-values
of H$_{2}$O$_{2}$ as well.\cite{shininvestigation2024} This behavior
was recently studied in simulations by Shin \emph{et al. }\cite{shininvestigation2024}
which showed by \emph{Topas-nBio} based MCS, that the ratio of the
hydrogen peroxide yields between LDR/UHDR changes in dependence of
the scavenging capacity of an additional organic hydroxyl radical
scavenger. The change from $\frac{g_{LDR}(H_{2}O_{2})}{g_{HDR}(H_{2}O_{2})}<1$
to $\frac{g_{LDR}(H_{2}O_{2})}{g_{HDR}(H_{2}O_{2})}>1$ was predicted
to happen at a scavenging capacity of around (10$^{3}$-10$^{4}$)$\,s^{-1}$.
This was simulated for 5.5$\,$MeV electrons, with an average DR of
0.1$\,$Gy/s for LDR, and 5.6$\times10{}^{6}$$\,$Gy/s for HDR, within
water containing 4$\,$\% O$_{2}$ and a generic hydroxyl radical
scavengers with scavenging capacities covering the range between (10$^{3}$-10$^{10}$)$\,s^{-1}$.
These results might already explain the observed differences between
the 'older' studies, which all were performed in glassware, while
the more 'recent' studies, which all were performed in various plastic
containers - mostly PCR tubes made out of polypropylene.\footnote{Still, it is worth noting here, that such carbon
contaminations may actually better represent the scavenging capacity
of a cellular environment than ultrapure water.\cite{shininvestigation2024}}\\ 
Additionally, similar effects were reported by the same authors for the prediction
of DR dependent oxygen consumption, where the addition of a 'generic organic
carbon', helped to reproduce the experimental data concerning oxygen depletion in the LDR regime
from Jansen \emph{et al.}\cite{jansendoes2021}. In this study the irradiation were performed in 3D printed sampleholders made from a proprietary organic compound called \emph{VeroClear} - hinting in the same direction. \\
It might be tempting to invoke Occam’s razor here, and to argue immediately, that the obvious reason of the discrepancy between the various studies are just caused by the presence of organic contaminants. 
However, correlation is not causation, therefore this assumption has to be thoroughly tested before taken as granted. Especially the release of microplastic \cite{braunsmart2021} from these PCR tubes and similar containers should be carefully quantified under standard incubation conditions, as well as when exposed to radiation, which may lead to
additional release of polymer fragments from the surface into the bulk water. Since such data is currently absent from the literature, it is worth to consider other possibilities as well. 
\subsection*{Slightly more complex: pH and assays}
In most of the studies ultrapure water without any buffering component
was used. Nominally ultrapure water has a pH of 7. However, this can
quickly vary due to the uptake of ambient CO$_{2}$ which forms carbonic
acid in water an naturally leads to a more acidic pH.\cite{zhanganalysis2024}
Similarly, radiation pulses themself can lead to the transient formation
of H$_{3}$O$^{+}$ and alter the pH.\cite{hahnsparing2025} Therefore
tight control or at least monitoring is a prerequisite to obtain reliable
results. Especially when we consider the strong pH dependence of the
H$_{2}$O$_{2}$ g-values as reported by various authors.\cite{sehestedrate1968,rotheffect2011}
Additionally the standard \emph{AmplexRed} assay has a pH dependence
and might be a source of uncertainty when used under in an environment
which is not tightly controlled for these fluctuations. Therefore
other fluorescence assays, which are suitable for usage over broader
pH range, might be a safer choice. Another quite obvious method would
be to measure the time dependent absorbance of H$_{2}$O$_{2}$ ($\lambda_{max}^{absorbance}$=200$\,$nm,
extinction: $\varepsilon$=189$\,$M$^{-1}$cm$^{-1}$), but that might
be complicated by the simultaneously overlapping absorption from other
reactive oxygen species (ROS) in this region.\cite{burnsmethods2012,buxtoncritical1988}
A comprehensive overview of alternative detection methods for hydrogen
peroxide, the hydroxyl radical, superoxide and singlet oxygen can
be found in the overview article by Burns \emph{et al.}\cite{burnsmethods2012},
while further specialized methods for H$_{2}$O$_{2}$ detection were
reviewed by Gulaboski \emph{et al.}\cite{gulaboskireview2019}
\section{The simulation side}
In addition to the experimental conditions, we must of course also
take the simulation models into account. Here, the simulation
parameters, the models themselves, and their validity ranges - which
may not longer apply under different dose rates - are possible sources
of error. Standard reaction kinetics can be applied to predict kinetics
in a deterministic manner for uniform concentrations of reactants
within a fixed volume. However, these assumptions do not hold for radiation
tracks.\cite{planteconsiderations2017} While, estimates based on
reaction kinetics can provide useful insights, more complex systems
with inhomogeneous distribution of reactants, or complex geometries
can be studied in more detail by MCS. Parameters influencing the outcome
of the calculations of reaction kinetics, MCS based on a step-by-step
(SBS) approach, or the independent-reaction time (IRT) method require
at least an appropriate set of possible reactions and rate constants,
and depending on the exact type of model additional values such as
diffusion constants, initial spatial distribution and concentrations
of the reactants, as well as their charge and spin configuration.\cite{planteconsiderations2017}
Even though the homogeneous chemistry models are overall well understood
and tested, for some conditions open questions still remain.
For example, Pastina and LaVerne measured the effects of added molecular
hydrogen on \ce{H2O2} yields during water radiolysis with radiation
of varying LET. They observed differences between the escape yields
of \ce{H2O2} between experiment and track-structure simulation, which
let them conclude, that the homogeneous models to study the long-time
chemistry of water cannot be applied to high LET radiation, and that
a relevant yield of oxidizing species might be produced. \cite{pastinaeffect2001}
Similar effects may be relevant for the UHDR case, where the higher
intertrack recombinations may behave somewhat similar to the increase
of intratrack combination when comparing low and high LET radiation.
Even more uncertain factors are introduced during the simulation of
the physico-chemical and non-homogeneous stages which will be discussed
in the following. 
\subsection*{Initial conditions and many particle interactions}
The initial spatial distribution and amount of excited species depends
strongly on the structure of the radiation track, and therefore on
the accuracy of the related elastic and inelastic scattering cross
sections, as well as the branching ratios for the products following
water radiolysis.\cite{greenstochastic1990,pimblottstochastic1991,hahnaccessing2023}
These values are particle, solvent and energy dependent. Values of
the reaction radii and the assumptions about Geminate recombination
of, for example the recombination of electron with the excited water
cation, influence the initial configuration of the system similarly.
This initial distribution of the particles within the spurs and tracks
is an essential prerequisite to conduct meaningful SBS or IRT based
MCS. Therfore the related underlying models and and input parameters
are worth to be reviewed thoroughly.\\
In the case of UHDR or very high LET, the radical production could be high
enough that the simplified pairwise interaction might become an oversimplification
and the reactions cannot be treated as isolated two-body problem anymore.
Similarly, high concentrations of charged radicals close to each other
can influence the local dielectric properties of water temporarily,
as it happens as well when salts or zwitterionic cosolutes are present.
\cite{kaidynamics2014,hahncombined2016} On short time scales, after
initial ionization events, this may have substantial effect on the
processes in the solvent itself, and as well on the types and rates of reactions
produced with biomolecules such as DNA and proteins.\cite{kaidynamics2014,kaidynamic2016} \\
IRT based simulations are often preferred over SBS approaches due
to their increased simulation speed, which allows to study extended
systems over an extended time. Even though, IRT based MCS consider
particle movement only implicitly, it was shown that IRT and SBS methods
are exactly equivalent for two-particle systems.\cite{planteconsiderations2017}
However, when multiple charged reactive species are produced close
to each other, this equivalence and the effects of many particle interactions
over short distances, as well as their interplay with the solvents
and other cosolutes have to be studies in more detail. If an evaluation
would show a related problem, a carefully adjusted transition from
short-term calculations by a suitable method and long term simulations
might be a solution here. A similar approach was already discussed
by the authors of the \emph{Topas-nBio} toolkit, for the transition
between IRT and the Gillespie formalism to optimize the simulation
of the long term chemistry after water radiolysis.\cite{d-kondointegrated2023}
\subsection*{Beyond diffusion controlled movements}
During SBS based MCS simulations only particle movements by Brownian
diffusion and reactions from particle-particle collisions are considered.\cite{planteconsiderations2017}
However, various studies predicted, that irradiations performed with
heavy particles such as ions, may produce 'shock waves', \cite{deverasimulation2018,deverathermomechanical2017,solovyovphysics2009}and
can lead to non-equilibrium conditions - these may alter the radical
transport completely and facilitate intertrack recombination substantially.\cite{hahnsparing2025}
Similarly, a spatially non homogeneous dose deposit distribution may
lead to a temperature gradient and provide an additional source for
non-diffusion base particle transport \emph{via} processes such as
convection.\cite{hahnmeasurements2017} These two mechanisms would
result in non-diffusion controlled reaction dynamics and aren't currently captured
at all by the MCS methods mentioned above. Here, molecular-dynamics
simulations with reactive force fields may provide an approach to
study such effects - with the backdrop of higher computational cost.\cite{deverathermomechanical2017,deverasimulation2018}
\section{Solving the riddle?}
We have aimed to understand the discrepancies between different experimental
and simulation studies in terms of the dose-rate dependent hydrogen
peroxide yield. This was motivated by the need for a coherent understanding
of the radiation chemistry underlying the studies of the radiotherapeutical
FLASH effect, before moving from simplified model systems, towards
more complex biomimetic or even living systems. Therefore we looked
at the differences and common denominators of these studies, and possible
explanations for the observed effects. On the experimental side, concrete
suggestions regarding the materials and purity of sample containers,
as well as with respect to the measurement methods of hydrogen peroxide yields were made. 
The implementation of these need careful and tight control of all experimental parameters, and a subsequent reporting
of all (!) of them, to allow for reproducibility and clarity.\cite{schulerultrahigh2022}
Otherwise an in depth understanding of the underlying processes will be difficult to obtain. 
In terms of the simulation the most relevant parameters were discussed and some factors such as the initial distribution of particles within a track and factors influencing their early recombinations were highlighted. 
Possible effects by non-diffusion controlled particle displacement, and changes in the interaction by modified dielectric
properties or many particle interactions may need be studied in much more detail by advanced molecular dynamics simulations. They can provide further inside into the applicability of Monte-Carlo based approaches. Such work should be closely accompanied by time-resolved spectroscopic studies of the very early time points in particle interactions, recombinations and reactions.
\section{Looking towards radiation biology}
Once these open questions have been clarified, it will be time to
take a step further and turn to systems that correspond to a more
realistic representation of cells, than ultrapure water. In cells,
a completely different chemical and scavenging environment is present
due to strictly controlled pH, high salt concentrations, interactions
with DNA, lipids, thiols, and proteins, and especially enzymes such
as superoxide dismutase (SOD), which affect the net radical yields
substantially.\cite{wardmanapproaches2022} Furthermore, the dielectric
properties of interfacial waters change compared to the bulk. Therefore
the screening of particle-particle interactions can be expected to
be modified as well close to interfaces as found in biomolecules -
which is the most relevant place for biologically relevant damage within cells. 
Additionally the molecular crowding and decreased viscosity within the cells affect and restrict particle
movement and diffusion, as discussed by Luby-Phelps and references
therein.\cite{luby-phelpsphysical2013}\\
Similar arguments apply to the experiments beyond the \ce{H2O2} yields, which lead to rejection
of the oxygen depletion theory.\cite{jansendoes2021,thomasproton2024} 
They were performed in artificial systems these are not representative
of conditions found within the cytosol, mitochondria or even the nucleus.
Since an oxygen gradient does not only exist in tissue, but as well
within a cell, a possible local depletion of oxygen might still be
relevant, when analyzed within a more realistic environment, representing
either mitochondria or a cell nucleus. There, the different oxygen
tensions, types of salts, the presence of SOD (mitocondria), or chromatin
and histones (nucleus), and other DNA binding and repair proteins
\cite{hallierradiation2025} may completely alter the radiation
sensitivity of the respective mitochondrial DNA and chromosomal DNA.\\
In conclusion, to make a successful leap from radiation chemistry
towards radiation biology, the fundamentals have to be solidly grounded
in a understanding of the underlying chemical mechanisms. In addition it
has to be taken care that the details and specific of the cellular
environment under study are represented in sufficient detail.
\backmatter
\bmhead{Acknowledgements}
Discussions with A.\,Solov'yov, R.\,Erdmann, and A.\,Adhikary are gratefully acknowledged. 
\section*{Declarations}
\subsection*{Funding}
This work was funded by the Deutsche Forschungsgemeinschaft (DFG, German Research Foundation) under grant number 442240902 (HA 8528/2-2).
\subsection*{Competing interests}
The authors declare no competing interests.
\section*{Author contributions}
M.B.\,Hahn wrote the manuscript and obtained the funding. 
\subsection*{Materials, Data and Correspondence}
No new data was generated to write this article.
Correspondence should be addressed to the corresponding author M.B.H. (email: marc-benjamin.hahn@fu-berlin.de).
\begin{appendices}
\end{appendices}


\end{document}